\documentclass[a4paper,11pt]{article}

\usepackage{amsfonts, amsmath, amssymb, amsthm}
\usepackage[colorlinks=true,citecolor=blue]{hyperref}
\usepackage[margin = 2.50cm]{geometry}
\usepackage{mathtools}
\usepackage{graphicx}
\usepackage{enumerate}
\usepackage{verbatim}
\usepackage{tikz}
\usepackage{algorithmic}
\usepackage[utf8]{inputenc}
\usepackage{microtype}
\usepackage{amsmath,amsthm,amssymb, color, tikz, mathtools}
\usepackage{lipsum,booktabs}
\usepackage{wrapfig}
\usepackage{tcolorbox}
\usepackage{mdframed}
\usepackage[utf8]{inputenc}
\usepackage{algorithm2e}
\usepackage[colorlinks=true,citecolor=blue]{hyperref}
\usepackage{caption}
\usepackage{subcaption}

\usetikzlibrary{positioning}
\usetikzlibrary{shapes}
\usetikzlibrary{decorations.pathreplacing}
\usepackage{enumerate, enumitem}

\usepackage{wrapfig}

\SetKwComment{Comment}{/*}{*/}

\newtheorem{theorem}{Theorem}[section]
\newtheorem{corollary}[theorem]{Corollary}

\newtheorem{lemma}[theorem]{Lemma}

\newtheorem{conjecture}[theorem]{Conjecture}

\newcommand{\NN}{\mathbb{N}}

\newcommand{\RR}{\mathbb{R}}
\newcommand{\R}{\mathbb{R}}

\newcommand{\cB}{\mathcal{B}}

\newcommand{\cF}{\mathcal{F}}
\newcommand{\cG}{\mathcal{G}}

\newcommand{\Soumi}[1]{\textcolor{cyan}{\textbf{Soumi:} #1}}

\newcommand{\remove}[1]{}
\newcommand{\dist}[3]{\mathrm{dist}_{#1}\left( #2, #3\right)}

\newcommand{\tc}{\textcolor}

\title{
Colorful two-piercing theorem for boxes
}

\author{
Sourav Chakraborty\footnote{Indian Statistical Institute, Kolkata, India}
\and 
Arijit Ghosh\footnotemark[1]
\and
Soumi Nandi\footnote{The Institute of Mathematical Sciences, Chennai, India}
}

\date{}

\begin{document}

\maketitle

\begin{abstract}
   We prove a colorful extension of a Helly-type theorem by Danzer and Grünbaum (\textit{Combinatorica}, 1982) concerning two-piercing families of axis-parallel boxes in $\mathbb{R}^d$. We also show that our result is tight by constructing extremal families that achieve the bound. Related work includes a graph-theoretic proof of the original theorem by Pendavingh, Puite, and Woeginger (\textit{Discrete Applied Mathematics}, 2008), and a two-piercing result for lower-dimensional boxes by Baños and Oliveros (\textit{Acta Mathematica Hungarica}, 2018).
\end{abstract}

\section{Introduction}

Let $S$ be a subset of $\mathbb{R}^d$ and $\mathcal{F}$ a family of subsets of $\mathbb{R}^d$. We say that $S$ {\em pierces/hits} $\mathcal{F}$ if for all $B \in \mathcal{F}$, we have $S \cap B \neq \emptyset$. For any natural number $n$, a family $\mathcal{F}$ is said to be {\em $n$-pierceable} if there exists a subset $A \subseteq \mathbb{R}^d$ with $|A| \leq n$ such that $A$ pierces $\mathcal{F}$. 
{\bf Unless stated otherwise, all families of geometric objects considered in this paper are finite.}

Helly's theorem~\cite{Helly23}, one of the fundamental results in discrete geometry, states that for any family $\mathcal{F}$ of convex sets in $\mathbb{R}^d$, if every $(d+1)$-tuple from $\mathcal{F}$ is one-pierceable, then the whole family $\mathcal{F}$ is one-pierceable. Over the years, ``Helly-type" theorems have been thoroughly studied and found various applications (see~\cite{AmentaDeLS17}). 

Katchalski and Liu~\cite{10.2307/2042758} generalized Helly's theorem by weakening its assumption. Instead of requiring all $(d+1)$-tuples to be one-pierceable, they showed that if a positive fraction of the $(d+1)$-tuples are one-pierceable, then there exists a subfamily of $\mathcal{F}$, with size at least a certain fraction of $|\mathcal{F}|$, that is one-pierceable. Specifically, for any $\alpha \in (0,1]$, there exists $\beta = \beta(\alpha, d) \in (0, 1]$ such that if at least $\alpha$ fraction of all $(d+1)$-tuples from $\mathcal{F}$ is one-pierceable, then there exists a subfamily of $\mathcal{F}$ of size at least $\beta |\mathcal{F}|$ that is one-pierceable. This result is known as the {\em fractional Helly theorem}.

Let $\mathcal{F}_1, \dots, \mathcal{F}n$ be non-empty families of convex sets in $\mathbb{R}^d$. A $t$-tuple $(C_1, \dots, C_t)$ is a {\em colorful $t$-tuple} from the above families of convex sets if for each $j \in [t]$, there exists $i_j \in [n]$ such that $C_j \in \mathcal{F}_{i_j}$, and for any pair ${j, k} \subseteq [t]$, we have $i_j \neq i_k$.
Lovász, and later Bárány~\cite{Barany82}, also introduced a fundamental generalization of Helly's theorem known as the {\em colorful Helly theorem}.
\begin{theorem}[Lov\'{a}sz and B\'{a}r\'{a}ny~\cite{Barany82}: Colorful Helly Theorem]\label{th:barany} 
If $\mathcal{F}_1, \mathcal{F}_2,\dots, \mathcal{F}_{d+1}$ are families of convex sets in $\mathbb{R}^d$ such that every colorful $(d+1)$-tuple is one pierceable, then at least one of the families $\mathcal{F}_i$ is one pierceable.
\end{theorem}

By applying the colorful Helly theorem, we can directly derive the original Helly's theorem. Additionally, the proof of Theorem~\ref{th:barany} by Bárány~\cite{Barany82} can be 
directly extended to prove the following result:

\begin{theorem}[B\'{a}r\'{a}ny~\cite{Barany82}: Strong Colorful Helly Theorem]\label{prop: strong_col_Helly}
Suppose $d$ and $n$ are natural numbers with $n \geq d+1$.
Let $\mathcal{F}_{1}, \mathcal{F}_{2}, \dots, \mathcal{F}_{n}$ be families of convex sets in $\mathbb{R}^d$ such that every colorful $(d+1)$-tuple is one pierceable. Then, there exists a subset $S$ of $[n]$ of size $n-d$ and for all $k \in [n] \setminus S$ there exists $F_k\in\mathcal{F}_k$ such that the extended family 
$$
    \left(\bigcup_{i \in S}\mathcal{F}_{i} \right) \bigcup \Big\{F_k\;\big|\: k\in [n]\setminus S \Big\}
$$ 
is one-pierceable.
\end{theorem}

Kalai and Meshulam~\cite{KALAI2005305} gave topological and matroidal extensions of the above colorful results. Holmsen and Lee~\cite{holmsen2019radon} showed that, in general, convexity spaces with bounded {\em Radon number} imply both a colorful Helly theorem and a fractional Helly theorem.

Danzer and Grünbaum~\cite{DBLP:journals/combinatorica/DanzerG82} explored the possibility of establishing a Helly-type theorem for $n$-pierceability in arbitrary families of convex sets. Unfortunately, for $n \geq 2$, they demonstrated that such a Helly-type theorem does not hold for arbitrary families of convex sets, see~\cite[page 17]{hadwiger1960}. 
Chakraborty et al.~\cite{DBLP:journals/corr/ChakrabortyPRS13} showed the impossibility of getting a fractional Helly theorem for $n$-pierceability for even closed balls in $\R^{2}$.
However, for special classes of convex sets, such as families of axis-parallel boxes in $\mathbb{R}^{d}$,
Danzer and Grünbaum~\cite{DBLP:journals/combinatorica/DanzerG82}
established results similar to Helly's theorem for $n$-pierceability. {\bf Note that, unless stated explicitly otherwise, the term {\em box/boxes} will refer to {\em axis-parallel box/boxes} for the rest of the paper.}

Before stating Danzer and Grünbaum's results, we introduce some notations. For natural numbers $d$ and $n$, let $h : = h(d,n)$ be the smallest natural number such that every family $\mathcal{F}$ of boxes in $\R^{d}$ is $n$-pierceable if every family of at most $h$ boxes from $\mathcal{F}$ is $n$-pierceable. 

Observe that $h(d,1) = 2$ directly follows from Helly's theorem, and it is a simple exercise to show $h(1, n) = n+1$.
Danzer and Grünbaum~\cite{DBLP:journals/combinatorica/DanzerG82} 
showed that for $d \geq 2$, $n \geq 3$ and $(d,n) \neq (2,3)$, 
$h(d,n) = \aleph_{0}$\footnote{For natural numbers $d$ and $n$, we define $h(d, n) = \aleph_{0}$ to mean that for every $k \in \NN$, there exists a family $\mathcal{G}_{k}$ of boxes in $\R^{d}$ such that any $k$ boxes from $\mathcal{G}_{k}$ can be pierced by $n$ points, but the entire family $\mathcal{G}_{k}$ cannot be pierced by $n$ points. This notation was introduced by Danzer and Grünbaum~\cite{DBLP:journals/combinatorica/DanzerG82}\remove{, and we adopt it for the rest of the paper}.}. For general $d$, Danzer and Grünbaum proved the following fundamental Helly-type theorem for $2$-pierceability of boxes.

\begin{theorem}[Danzer and Gr\"unbaum~\cite{DBLP:journals/combinatorica/DanzerG82}]\label{prop: Danzer}
For all natural numbers $d$, we have 
$$
    h(d,2) = \begin{cases}
    3d,      & \quad \text{if } d \text{ is odd}\\
    3d-1,  & \quad \text{if } d \text{ is even}
    \end{cases}
$$
\end{theorem}
\noindent
Additionally, Danzer and Gr\"unbaum~\cite{DBLP:journals/combinatorica/DanzerG82} proved that $h(2,3) = 16$. Pendavingh, Puite, and Woeginger~\cite{PendavinghPW08} later provided an elegant graph-theoretic proof of this result. More recently, Ba\~nos and Oliveros~\cite{BanosOliveros2018} investigated $2$-piercings of families of boxes in $d$-dimensional Euclidean space by reducing the dimension of the boxes relative to the ambient space.

\subsection*{Our results}

Suppose $H_c(d,2)$ is the smallest positive integer such that the following holds:
If \( \mathcal{F}_1, \mathcal{F}_{2}, \dots, \mathcal{F}_{H_c} \) are families of boxes in \( \mathbb{R}^d \) such that every colorful \( H_c \)-tuple \( (F_1, F_{2}, \dots, F_{H_c}) \) is \( 2 \)-pierceable, then there exists an index \( i \in [H_c] \) such that for each \( k \in [H_c] \setminus \{i\} \), there exists \( F_k \in \mathcal{F}_k \) for which the extended family  
\[
    \mathcal{F}_i \cup \{ F_k \mid k \in [H_c] \setminus \{i\} \}
\]
is \( 2 \)-pierceable.



The main contribution of this paper is the following colorful generalization of the $2$-piercing theorem for boxes due to Danzer and Gr\"{u}nbaum (Theorem~\ref{prop: Danzer}).
\begin{theorem}[Colorful Helly theorem for $2$-piercing for boxes]\label{th: main strong theorem}
    For all natural numbers $d$, we have $H_{c}(d,2) = 3d$.
\end{theorem}

One of the interesting things to observe is that, unlike other colorful Helly-type theorems, Theorem~\ref{th: main strong theorem} is not a generalization of Danzer and Gr\"unbaum's $2$-pierceability theorem for boxes (Theorem~\ref{prop: Danzer}). That is, we cannot derive Theorem~\ref{prop: Danzer} from Theorem~\ref{th: main strong theorem}. In Section~\ref{sec: h(d,2)} we prove $H_{c}(d,2) \leq 3d$, and in Section~\ref{sec: extremal constructions} we show that $H_{c}(d,2) > 3d-1$.

Observe that, surprisingly, when \( d \) is even, \( H_c(d,2) > h(d,2) \). Since Theorem~\ref{th: main strong theorem} proves a stronger variant of the \emph{colorful Helly Theorem}, it is natural to expect that the following statement is true:
\begin{center}
    \emph{Let \( d \in \mathbb{N} \) be an even number, and let \( \mathcal{F}_1,\mathcal{F}_{2}, \dots, \mathcal{F}_{3d-1} \) be families
    of boxes in \( \mathbb{R}^d \).\\
    If every colorful \( (3d-1) \)-tuple is \( 2 \)-pierceable, then
    there exists \( i \in [3d-1] \)\\
    such that \( \mathcal{F}_i \) is \( 2 \)-pierceable.}
\end{center}
\noindent
We provide explicit families of boxes in \( \mathbb{R}^d \) that disproves the above statement.

\begin{theorem}
   \label{theorem:extremalconstructions}
            For every $d\in\mathbb{N}$, there exist non-empty families $\mathcal{F}_{1}, \mathcal{F}_{2}, \dots, \, \mathcal{F}_{3d-1}$ of \remove{axis-parallel }boxes in $\R^d$ such that 
            \begin{itemize}
                \item 
                    every colorful $(3d-1)$-tuple is $2$-pierceable, and 
    
                \item
                    for each $i \in [3d-1]$, $\mathcal{F}_{i}$ is not $2$-pierceable.
            \end{itemize}
\label{Extremal constructions}
\end{theorem}

\noindent

\section{Notations} 

We will use the following notations in this paper. 
\begin{enumerate}
    \item 
        For all $n \in \mathbb{N}$,
        $[n] := \left\{ 1, \, \dots, \, n\right\}$. 
        By $[0]$, we mean the empty set.
        
    \item
        For all $u, \, v \in \mathbb{R}^{d}$, $\langle u,v\rangle$ denotes the {\em inner product} between $u$ and $v$.
        
    \item
        For any set $S$ with $|S| \geq k$, $\binom{S}{k}$ denotes the family of all $k$-sized subsets of $S$.
        
    \item
        For all $i \in [d]$, we denote by $e_{i}$ the unit vector whose $i$-th co-ordinate is $1$ and the rest of the co-ordinates are $0$. 
        
    \item
        For any box $B = [\alpha_{1}, \beta_{1}]\times [\alpha_{2}, \beta_{2}] \times \dots \times [\alpha_{d}, \beta_{d}]$ in $\R^d$ and $j \in [d]$, $\pi_{j}(B)$ denotes projection of $B$ on the $j$-th coordinate axis, that is, 
        the interval $I=[\alpha_{j},\beta_{j}]$.
        
    \item
        For any two boxes $B, B'\in \R^d$, if $\pi_{j}(B)=[\alpha,\beta]$ and $\pi_{j}(B') = [\alpha',\beta']$ such that $\beta<\alpha'$ then we denote the distance between $B$ and $B'$ along $j$-th coordinate axis by $\dist{j}{B}{B'} =\alpha'-\beta$.

    \item
        Let $B = [\alpha_{1}, \beta_{1}] \times [\alpha_{2}, \beta_{2}] \times \dots \times [\alpha_{d}, \beta_{d}]$ be\remove{an axis parallel} a box. We say $B'$ is a {\em face} of $B$ if there exists a $j \in [d]$ such that $B'$ is either this
        $$ 
            [\alpha_{1}, \beta_{1}]\times \dots \times [\alpha_{j-1}, \beta_{j-1}]\times \{\alpha_{j}\} \times [\alpha_{j+1}, \beta_{j+1}]\times \dots \times [\alpha_{d}, \beta_{d}]
        $$
        or
        $$ 
            [\alpha_{1}, \beta_{1}]\times \dots \times [\alpha_{j-1}, \beta_{j-1}]\times \{\beta_{j}\} \times [\alpha_{j+1}, \beta_{j+1}]\times \dots \times [\alpha_{d}, \beta_{d}].
        $$ 
    
         \item
        Let $B = [\alpha_{1}, \beta_{1}] \times [\alpha_{2}, \beta_{2}] \times \dots \times [\alpha_{d}, \beta_{d}]$ be a\remove{n axis parallel} box. We say $\lambda=(\lambda_1, \dots,\lambda_d)\in\R^d$ is a {\em vertex} of $B$ if for each $j\in [d]$, $\lambda_j\in\{\alpha_j,\beta_j\}$
        
        \item
            Let $B = [\alpha_{1}, \beta_{1}] \times [\alpha_{2}, \beta_{2}] \times \dots \times [\alpha_{d}, \beta_{d}]$ be a\remove{n axis parallel} box. A pair of vertices $(p, q)$ in $B$ is called {\em diagonally opposite} if $p = (\lambda_{1}, \dots,  \lambda_{d})$, $q = (\eta_{1}, \dots,  \eta_{d})$ and $\forall j\in [d], \{\lambda_j, \eta_j\}=\{\alpha_j,\beta_j\}$. Clearly, 
            for every vertex $p\in B$, there is a unique vertex $q\in B$ such that $(p,q)$ is diagonally opposite.
        
       \item Let $I=[\alpha,\beta]$ and $I'=[\alpha',\beta']$ be two intervals on ${\R}$. We write $I<I'$ if and only if $\beta<\alpha'$.

\end{enumerate}

\section{Colorful Helly Theorem for two-piercing boxes in $\mathbb{R}^{d}$}
\label{sec: h(d,2)}

In this section, we will prove the following result.  

\begin{theorem}[Upper bound part of Theorem~\ref{th: main strong theorem}\remove{ for $n=2$ and general $d$}]
\label{thm:restatement_strong-h(d,2)}
    For all $d \in \mathbb{N}$, we have $H_c(d,2)\leq 3d$.
\end{theorem}

In Section~\ref{ssec:properties-axis-parallel-boxes}, we establish key properties of boxes required for the proof of Theorem~\ref{thm:restatement_strong-h(d,2)}, which is provided in Section~\ref{ssec:proof-restatement-h(d,2)}.

\subsection{Properties of \remove{axis-parallel }boxes}
\label{ssec:properties-axis-parallel-boxes}

Using Theorem~\ref{prop: Danzer} and Theorem~\ref{prop: strong_col_Helly} we get the following corollary. 

\begin{corollary}\label{lem: every_col_pair_intersects}
Let $\mathcal{F}_{1}, \mathcal{F}_{2}, \dots, \mathcal{F}_{n}$ be families of boxes in $\R^{d}$, with $ n \geq d+1$. Suppose every colorful pair of boxes intersects (that is, every pair consisting of one box from \( \mathcal{F}_i \) and one from \( \mathcal{F}_j \), with \( i \neq j \), has non-empty intersection). Then there exists a subset \( S \subseteq [n] \) of size \( n - d \) such that for all \( k \in [n] \setminus S \), there exists a box \( B_k \in \mathcal{F}_k \) for which the extended family
\[
    \left( \bigcup_{i \in S} \mathcal{F}_i \right) \cup \left\{ B_k \mid k \in [n] \setminus S \right\}
\]
is one-pierceable.
\end{corollary}

\begin{proof}
    Since every colorful pair of boxes has a non-empty intersection, using Theorem~\ref{prop: Danzer}~(a), we get that every colorful $(d+1)$-tuple 
    is one-pierceable. Rest follows directly from Theorem~\ref{prop: strong_col_Helly}.
\end{proof}

\begin{lemma}\label{lem: pierceable by a hyper plane}
Let $\mathcal{F}_{1}, \mathcal{F}_{2}, \dots, \mathcal{F}_{n}$ be families of boxes in $\R^{d}$, with $ n \geq d+1$.
Suppose there exists $j \in [d]$ such that for all $B \in \mathcal{F_{\ell}}$ and $B' \in \mathcal{F}_{i}$, with $\ell \neq i$,  we have $\pi_{j}(B)\cap \pi_{j}(B')\neq \emptyset$. Then there exists a subset $S \subseteq [n]$ with $|S| = n-1$ and there exists $B_k\in\cF_k$, with $\{k\}=[n]\setminus S$, such that the family 
\begin{align*}
    \left( \bigcup_{i\in S}\mathcal{F}_{i} \right) \cup \big\{ B_{k} \big\}
\end{align*}
can be pierced by a single hyperplane orthogonal to the $j$-th coordinate axis.
\end{lemma}

\begin{proof}
    For every $i\in [n]$, we define 
    $\mathcal{G}_{i}:= \left\{ \pi_{j}(B)\,\mid\, B\in \mathcal{F}_i \right\}$. Observe that every colorful pair of intervals from $\mathcal{G}_{1}, \dots, \mathcal{G}_{n}$ have a non-empty intersection. Therefore, the result now directly follows by applying Theorem~\ref{prop: strong_col_Helly} to the families $\mathcal{G}_{1}, \dots, \mathcal{G}_{n}$. 
\end{proof}
    
The proof of the following result appears in Danzer and Grünbaum~\cite{DBLP:journals/combinatorica/DanzerG82} (see the last two paragraphs of page 239).

\begin{lemma}
    Let $V=[\alpha_1,\beta_1]\times[\alpha_{2},\beta_{2}]\times\dots\times[\alpha_d,\beta_d]$ be a\remove{n axis parallel} box in $\R^d$ and $\mathcal{G}$ be a family of boxes in $\R^d$ such that the following two conditions are satisfied:
\begin{enumerate}[label=(\roman*)]
    \item 
        \label{cond: contains_a_vertex}for any $B\in\mathcal{G}$, we have $\forall j\in [d], \;\pi_j(B)\cap \{\alpha_j,\beta_j\}\neq\emptyset$,
    that is, $B$ contains at least one vertex of $V,$ and
    
    \item \label{cond: hit_by_diag}
        $\mathcal{G}$ is pierceable by some\remove{ diagonally opposite pair of} vertices of $V$.
\end{enumerate}
Let $f(\cG)$ denote the number of pairs of opposite vertices of $V$ each of which pierces $\cG$. Then $f(\cG)$ is either $0$ or a power of $2$. 
\label{lem: DanzerG}\end{lemma}

Using the above result, we can prove the following.

\begin{lemma}\label{lem: reducing_choice_of_diag}
Let $V=[\alpha_1,\beta_1]\times [\alpha_{2}, \beta_{2}]\times\dots\times[\alpha_d,\beta_d]$ be a\remove{n axis parallel} box in $\R^d$ and $\mathcal{G}_{1}, \mathcal{G}_{2}, \dots,\mathcal{G}_n$ be families of \remove{axis-parallel }boxes in $\R^d$ such that the following two conditions are satisfied:
\begin{enumerate}[label=(\roman*)]
    \item 
        \label{cond: contains_a_vertex}for any $i\in [n]$ and any $B\in\mathcal{G}_i$, we have $\forall j\in [d], \;\pi_j(B)\cap \{\alpha_j,\beta_j\}\neq\emptyset$,
    that is, $B$ contains at least one vertex of $V,$ and
    
    \item \label{cond: hit_by_diag}
        every colorful $n$-tuple from $\mathcal{G}_{1}, \mathcal{G}_{2}, \dots,\mathcal{G}_n$ is pierceable by some diagonally opposite pair of vertices of $V$.
\end{enumerate}
If $(B_{1}, B_{2}, \dots, B_r)$, with $r<n$, is a colorful $r$-tuple such that there are at most $2^k$, with $k<d$, distinct diagonally opposite pairs of vertices of $V$, each of which pierce $(B_{1}, B_{2}, \dots, B_r)$, then one of the following two options must hold:
\begin{enumerate}[label=\upshape(\Roman*)]
    \item 
    \label{result: 2pierceable} If $(\lambda, \lambda')$ is any diagonally opposite pair of vertices of $V$ piercing $(B_1, B_{2}, \dots, B_r)$ and $\mathcal{G}_i$ is a family such that $\forall \ell\in [r]$ we have $B_\ell\not \in\mathcal{G}_i$ then $\mathcal{G}_i$ is pierceable by $(\lambda, \lambda')$.
    
    \item\label{result: existance_non-2pierceable} There exists $i\in [n]$, and there exists $B_{r+1}\in\mathcal{G}_i$, such that there are at most $2^{k-1}$ distinct diagonally opposite pair of vertices of $V,$ each of which pierces the colorful tuple  $(B_1, B_{2}, \dots, B_r, B_{r+1})$. 
\end{enumerate}

\end{lemma}

\begin{proof}
If $k=0$, that is, $(B_1, B_{2}, \dots, B_r)$ is pierceable by a unique diagonally opposite pair of vertices of $V$, then Option~\ref{result: 2pierceable} must be true.
For otherwise, we get a colorful $(r+1)$-tuple which is not pierceable by any diagonally opposite pair of vertices of $V$, a contradiction to Condition~\ref{cond: hit_by_diag}.

Now let us consider the case when $k>0$.
If  Option~\ref{result: 2pierceable} is true then there is nothing to prove.
Otherwise, there exists a diagonally opposite pair of vertices $(\lambda, \lambda')$ of $V$ piercing $(B_{1}, B_{2}, \dots, B_r)$ and there exists $i\in [n]$ such that 
\begin{itemize}
    \item 
        $\forall \ell\in [r]$ we have $B_\ell\not \in\mathcal{G}_i$, and
    
    \item 
        $\mathcal{G}_i$ is not piercable by $(\lambda, \lambda')$.
\end{itemize}
Thus, there exists $B_{r+1}\in\mathcal{G}_i$ such that $B_{r+1}\cap \{\lambda,\lambda'\}=\emptyset$. Then, by Lemma~\ref{lem: DanzerG}, there are at most $2^{k-1}$ diagonally opposite pairs of vertices in $V$, each of which pierces $(B_{1}, B_{2}, \dots, B_{r}, B_{r+1})$.
\end{proof}

\subsection{Proof of Theorem~\ref{thm:restatement_strong-h(d,2)}}
\label{ssec:proof-restatement-h(d,2)}

Let \(\mathcal{F}_1, \dots, \mathcal{F}_{3d}\) be non-empty families of boxes in \(\mathbb{R}^d\), with the property that every colorful \(3d\)-tuple is \(2\)-pierceable. We prove that there exists an index \(i \in [3d]\) such that for all \(k \in [3d] \setminus \{i\}\), there exists \(F_k \in \mathcal{F}_k\) such that  
$$
\mathcal{F}_i \cup \big\{ F_k \mid k \in [3d] \setminus \{i\} \big\}
$$ 
is \(2\)-pierceable. This establishes that \(H_c(d, 2) \leq 3d\) for all \(d \in \mathbb{N}\).

Observe that if every colorful pair of boxes from \(\mathcal{F}_1, \dots, \mathcal{F}_{3d}\) intersect, then by Corollary~\ref{lem: every_col_pair_intersects}, we get that there exists a subset $S$ of $[3d]$ of size $2d$ and for all $k \in [3d] \setminus S$ there exists $B_k\in\mathcal{F}_k$ such that the following extended family 
$$
    \left(\bigcup_{i \in S}\mathcal{F}_{i} \right) \cup \Big\{B_k\;|\: k\in [3d]\setminus S \Big\}
$$ 
is one-pierceable and we are done.

Therefore from now on we assume that there is at least one non-intersecting colorful pair of boxes.
Then there is at least one non-intersecting colorful pair of boxes, say $B_{1}, B_{2}$ and there exists $j\in [d]$ such that $\pi_j(B_1)\cap \pi_j(B_2)=\emptyset$. Let 
$$
    (B_{a}, B_{b})= \mathrm{argmax}\{\dist{j}{B}{B'}\; | \;(B,B')\text{ is a colorful pair  }\}. 
$$ 
Take $J=\{j\}$ and $I=\{i_{1}, i_{2}\}$, where $B_a \in \mathcal{F}_{i_1}$ and $B_b\in \mathcal{F}_{i_2}$. Suppose $\pi_j(B_a)=[\alpha_{j_1},\beta_{j_1}]$ and $\pi_j(B_b)=[\alpha_{j_2},\beta_{j_2}]$ and $\beta_{j_1}<\alpha_{j_2}$. Then for any $B\in\cF_i$ such that $i\in [3d]\setminus\{i_1, i_2\}$, we have $\pi_j(B)\cap\{\beta_{j_1},\alpha_{j_2}\}\neq\emptyset$.

Now suppose $r$ $(>0)$ is the largest integer such that we have the following: $\exists J = \left\{ j_{1} , \dots , j_{r} \right\} \subseteq [d]$ and $\exists I \remove{= \left\{ i_{1}, \dots, i_{2r} \right\}} \subseteq [3d]$ such that for each $i \in I$ there exists $B_{i} \in \mathcal{F}_i$ satisfying the following conditions:\begin{enumerate}[label=(\roman*)]
  
    \item\label{Condition (1)}  For all $j\in J$, there exists $j_1,j_2\in I$ such that $\pi_{j}(B_{j_{1}})=[\alpha_{j_{1}},\beta_{j_{1}}]$ and $\pi_{j}(B_{j_{2}})=[\alpha_{j_{2}},\beta_{j_{2}}]$ and $\beta_{j_{1}}<\alpha_{j_{2}}$. \label{cond: intersecting pair} 

    
\color{black}    
    \item \label{Condition (2)} For all $j \in J$ and for any $B \in \cF_{m}$, where $m \in [3d]\setminus \{j_1, j_2\}$, we have 
    $\pi_{j}(B) \cap \left\{ \alpha_{j_{2}}, \beta_{j_{1}} \right\}\neq\emptyset$.

    
\end{enumerate}
Since $r$ is the largest integer satisfying the above two conditions we can show that the following condition also holds: 
\begin{enumerate}
     \item[(iii)]\label{Condition (3)} for any $B\in\mathcal{F}_m$ and $B'\in\mathcal{F}_n$ such that $m\neq n $ and for any $j\in [d]\setminus J$, we have $\pi_j(B)\cap\pi_j(B')\neq\emptyset$.
\label{cond: col_proj_meets}\end{enumerate}
Now we only need to handle the following two cases:
\begin{itemize}
    \item
        Case 1: $r=d$, and 
    
    \item
        Case 2: $r<d$.
\end{itemize}

\paragraph{Case 1: $r=d$.} In this case $J=[d]$. Without loss of generality assume that $I=[p]$. 
Therefore $(B_1,\dots, B_{p})$, as a colorful tuple, must be $2$-pierceable. We claim that $(B_1,\dots, B_{p})$ is pierceable by at least one diagonally opposite pair of vertices of the box $D : = \prod\limits_{j\in[d]}[\beta_{j_1},\alpha_{j_2}]$. 
Now for each $j\in [d]$, we have, $B_{j_1}\cap B_{j_2}=\emptyset$. Without loss of generality, we assume that $B_{1_1}=B_1$ and $B_{1_2}=B_2$. Since $(B_1,\dots, B_{p})$ is $2$-pierceable, 
for each $j\geq 2$, there exists $B'_j\in\{B_{j_1}, B_{j_2}\}$ that intersects $B_{1}$ and there exists $B''_j\in\{B_{j_1}, B_{j_2}\}$ that intersects $B_{2}$ and moreover, we can choose $B'_j$ and $B''_j$ so that $\{B'_j,B''_j\}=\{B_{j_1}, B_{j_2}\}$ (for otherwise, $(B_1, B_2, B_{j_1}, B_{j_2})$ would not be $2$-pierceable, a contradiction). Again $2$-pierceability of $(B_1,\dots, B_{p})$ implies that \\

\begin{equation}
    \left(\bigcap_{j=2}^{d} B'_{j} \right)\bigcap B_1\neq\emptyset \quad \text{and} \quad \left( \bigcap_{j=2}^d B''_{j} \right)\bigcap B_2\neq\emptyset.
\label{eq: partition}\end{equation}


Note that, $\{B_1, B'_2,\dots,B'_d\},\;\{B_2, B''_2,\dots,B''_d\}$ are multisets of $\{B_1,\dots,B_p\}$ such that
$$
    \left\{B_1, B'_2,\dots,B'_d \right\} \cup \left\{B_2, B''_2,\dots,B''_d \right\} = \left\{ B_1, \dots, B_p \right\}.
$$
Additionally, observe that from Conditions~\ref{Condition (1)} and \ref{Condition (2)}, for all $j\in \left\{ 2, \dots, d \right\}$, we have
\begin{itemize}
    \item [(a)]
        $\pi_{j}(B_{j_1}) \cap \pi_{j}(B_{j_2}) = \emptyset$,

    \item [(b)]
        $\left|\left\{ \beta_{j_1}, \alpha_{j_2}\right\} \cap \pi_{j}\left( B_{j_1}\right) \right| = \left|\left\{ \beta_{j_1}, \alpha_{j_2}\right\} \cap \pi_{j}\left( B_{j_2}\right) \right| = 1$ and

    \item [(c)]
        $\left\{ \beta_{j_1}, \alpha_{j_2}\right\} \cap \pi_{j}\left( B_{1}\right)\neq \emptyset$ and $\left\{ \beta_{j_1}, \alpha_{j_2}\right\} \cap \pi_{j}\left( B_{2}\right)\neq\emptyset$.
    
\end{itemize}
For each $j\in\{2,\dots,d\}$, suppose $\lambda_{j} = \pi_j(B'_j) \cap \left\{ \beta_{j_1}, \alpha_{j_2} \right\}$ and $\lambda'_j=\pi_j(B''_j)\cap \left\{ \beta_{j_1}, \alpha_{j_2} \right\}$ such that $\{\lambda_j,\lambda'_j\}=\left\{ \beta_{j_1}, \alpha_{j_2} \right\}$.\remove{ Then $\dist{j}{B'_j}{B''_j}=\dist{}{\lambda_j}{\lambda'_j}$.} We claim that $\lambda_j\in\pi_j(B_1)$ and $\lambda'_j\in\pi_j(B_2)$. If possible let $\lambda_j\not\in\pi_j(B_1)$. Then by Property~\textcolor{red}{(c)}, $\lambda'_j\in\pi_j(B_1)$. Again by equation~(\ref{eq: partition}), $\pi_j(B_1)\cap\pi_j(B'_j)\neq\emptyset$. Since $\lambda_j$ is the end point of the interval $\pi_j(B'_j)$ nearest to $\lambda'_j$, so $\lambda'_j\in\pi_j(B_1)$ and $\pi_j(B_1)\cap\pi_j(B'_j)\neq\emptyset$ imply that $\lambda_j\in\pi_j(B_1)$, a contradiction.\remove{Then $\dist{j}{B_1}{B''_j}>\dist{}{\lambda_j}{\lambda'_j}=\dist{j}{B'_j}{B''_j}$, which contradicts Condition~\ref{cond: argmax}.} So $\lambda_j\in\pi_j(B_1)$. Similarly, we can show that $\lambda'_j\in\pi_j(B_2)$.

\color{black}
Now consider the vertex pair $(\lambda, \lambda')$ of $D$, where $\lambda :=(\beta_{1_1},\lambda_2,\dots,\lambda_d)$ and $\lambda' :=(\alpha_{1_2},\lambda'_2,\dots,\lambda'_d)$. Observe that  
\begin{itemize}
    \item $\lambda$ pierces $\{B_1,B'_2,\dots,B'_d\}$ and 
    \item $\lambda'$ pierces $\{B_2,B''_2,\dots,B''_d\}$.
\end{itemize}

We will now show that for all  $\tilde{B}_{i} \in \cF_{i}$, where $i > p$, the colorful $3d$-tuple 
$$
    \big( B_{1}, \dots, B_{p}, \tilde{B}_{p+1}, \dots, \tilde{B}_{3d} \big)
$$ 
is two-pierceable by some diagonally opposite pair of vertices from $D$. Since 
$$
    \big( B_{1}, \dots, B_{p}, \tilde{B}_{p+1}, \dots, \tilde{B}_{3d} \big)
$$
is two-pierceable (a precondition in the theorem) and for all $j \in [d]$ we have 
$B_{j_1} \cap B_{j_2} = \emptyset$, therefore the sets $[p]$ and $[3d]\setminus [p]$ can be written as 
\begin{itemize}
    \item
        $[p] = I' \cup J'$ such that $1\in I'$, $2\in J'$ and 
        
    \item
        $[3d]\setminus [p] = K'\cup L'$, with $K'\cap L' = \emptyset$
        \footnote{Note that $K'$ and $L'$ may not form a partition of $[3d]\setminus [p]$ as one of them can be an empty set. If $K' = \emptyset$ then by abuse of notation  $\left( \bigcap_{i \in I'} B_{i}  \right)\bigcap \left( \bigcap_{k \in K'} \tilde{B}_{k}\right)$ will mean the set $\bigcap_{i \in I'} B_{i}$, and similarly if $L' = \emptyset$ then $\left( \bigcap_{j \in J'} B_{j}  \right)\bigcap \left( \bigcap_{\ell \in L'} \tilde{B}_{\ell}\right)$ will mean the set $\bigcap_{j \in J'} B_{j}$.}
\end{itemize}
and satisfy the following:
\begin{align*}
    \left( \bigcap_{i \in I'} B_{i}  \right)\bigcap \left( \bigcap_{k \in K'} \tilde{B}_{k}\right) \neq \emptyset \quad \mbox{and} \quad
    \left( \bigcap_{j \in J'} B_{j}  \right)\bigcap \left( \bigcap_{\ell \in L'} \tilde{B}_{\ell}\right) \neq \emptyset.
\end{align*}
Again, using Conditions~\ref{Condition (1)} and \ref{Condition (2)}, we can show that there exists diagonally opposite pair of vertices $(\mu, \mu')$ of $D$ such that
$\mu \in \left( \bigcap_{i \in I'} B_{i}  \right)\bigcap \left( \bigcap_{k \in K'} \tilde{B}_{k}\right)$ and $\mu' \in \left( \bigcap_{j \in J'} B_{j}  \right)\bigcap \left( \bigcap_{\ell \in L'} \tilde{B}_{\ell}\right)$.

Now there are at most $2^{d-1}$ pair of diagonally opposite pair of vertices of $D$, such that each pair pierce $(B_1,\dots,B_{p})$. For each $i\in\{p+1,\dots, 3d\}$, let $\widetilde{\cF}_i=\cF_i\setminus\{B_1,\dots,B_{p}\}$. If $\widetilde{\cF}_i=\emptyset$, for some $i$, then any diagonally opposite pair of vertices of $D$ that pierce $(B_1,\dots,B_{p})$, also pierce $\widetilde{\cF}_i$. We will be done with the proof of {\bf Case 1} if this is the case since $\cF_{i} \subseteq \left\{ B_{1}, \dots, B_{p}\right\}$ and we have already shown that for all $\tilde{B}_{i}$, where $i \in [3d]\setminus [p]$, the colorful $3d$-tuple $(B_{1}, \dots, B_{p}, \tilde{B}_{p+1}, \dots, \tilde{B}_{3d})$ is two-pierceable by at least one diagonally opposite pair of vertices of $D$. 
Therefore, we can assume that for all $i\in\{p+1,\dots, 3d\}$, we have $\widetilde{\cF}_i\neq\emptyset$. 
Applying Lemma~\ref{lem: reducing_choice_of_diag} to the families $\widetilde{\cF}_{p+1}, \dots, \widetilde{\cF}_{3d}$, the axis-parallel box $D$, and the colorful $p$-tuple $(B_{1}, \dots, B_{p})$ we get that, for each $i \in [3d]\setminus [p]$, 
at least one of the following statements has to be true:
\begin{enumerate}[label=\upshape(\Roman*)]
    \item 
        For all diagonally opposite pair of vertices $(\lambda, \lambda')$ of $D$ piercing $(B_1,\dots, B_{p})$ and for all
        $i \in [3d]\setminus [p]$ we have $\widetilde{F}_{i}$ is also pierceable by $(\lambda, \lambda')$. Note that if $(\lambda, \lambda')$ pierces $(B_1,\dots, B_{p})$ and $\widetilde{F}_{i}$ then it also pierces $\cF_{i}$ as $\widetilde{\cF}_{i} \subseteq \cF_{i} \setminus \{ B_{1}, \dots, B_{p}\}$.

    \item
        There exists $\widetilde{B}_{i} \in \widetilde{\mathcal{F}}_i$ such that there are at most $2^{d-2}$ distinct diagonally opposite pair of vertices of $D$ each of which pierces the colorful tuple  $(B_1,\dots, B_p, \widetilde{B}_{i})$. 
\end{enumerate}

The above discussion shows that if we use the 
fact that for all $\widetilde{B}_{i} \in {\cF}_{i}$, with $i \in [3d]\setminus [p]$, the colorful $3d$-tuple $(B_{1}, \dots, B_{p}, \widetilde{B}_{p+1}, \dots, \widetilde{B}_{3d})$ is two-pierceable by a diagonally opposite pair of vertices of $D$, and repeatedly apply Lemma~\ref{lem: reducing_choice_of_diag} on the families ${\cF}_{p+1}, \dots, {\cF}_{3d}$ we get the following:
\begin{itemize}
    \item 
        There exists a colorful $(p+q)$-tuple $(B_1,\dots, B_{p}, \widetilde{B}_{i_{1}}, \dots, \widetilde{B}_{i_q})$, with $\widetilde{B}_{i_{j}} \in {F}_{i_{j}}$, 
        such that every family $\mathcal{F}_{i}$, for all $i \in [3d] \setminus \left([p]\cup \{i_{1}, \dots, i_{q}\}\right)$ is pierceable by a diagonally opposite pair of vertices $(\lambda, \lambda')$ of $D$, if $(\lambda, \lambda')$ also pierces $(B_1,\dots, B_{p}, \widetilde{B}_{i_{1}}, \dots, \widetilde{B}_{i_q})$.
\end{itemize}
Without loss of generality, assume that $3d \not\in [p]\cup \{i_{1}, \dots, i_{q}\}$. To complete the proof of {\bf Case 1}, 
take any $\widehat{B}_{i} \in \cF_{i}$, where $i \in [3d-1] \setminus \left([p]\cup \{i_{1}, \dots, i_{q}\}\right)$, and observe that there exists a diagonally opposite pair of vertices $(\lambda, \lambda')$ of $D$ such that $(\lambda, \lambda')$ pierces the extended family 
$$
    \cF_{3d} \cup \big\{ B_{i} \,\mid\, i \in [p] \big\} \cup \big\{ \widetilde{B}_{i_{j}} \,\mid\, j \in [q] \big\} \cup
    \big\{ \widehat{B}_{i} \, \mid \, i \in [3d-1] \setminus \left([p]\cup \{i_{1}, \dots, i_{q}\}\right) \big\}\footnote{Note that if $q = 0$, then the extended family will be
    $\cF_{3d} \cup \big\{ B_{i} \,\mid\, i \in [p] \big\} \cup
    \big\{ \widehat{B}_{i} \, \mid \, i \in [3d-1]\setminus [p] \big\}$.
    }.
$$

\paragraph{Case 2: $r<d$.} Without loss of generality, we assume that, $J=[r]$ and $I=[p']$. 
Then by Condition~\tc{red}{(iii)}\remove{~\ref{Condition (3)}(\Soumi{reference is not correct})}, for each $j\in \{r+1,\dots, d\}$ and for any $B\in\cF_m,\; B'\in\cF_n$ such that $m\neq n$, we have $\pi_j(B)\cap\pi_j(B')\neq\emptyset$. Then by Lemma~\ref{lem: pierceable by a hyper plane}, we get the following: for each $j\in \{r+1,\dots, d\}$ there exists $i_j\in [3d]$ and there exists $\hat{B}_{i_j}\in\cF_{i_j}$ such that $\{\hat{B}_{i_j}\}\cup\big(\bigcup_{k\in[3d]\setminus\{i_j\}}\cF_k\big)$ is pierceable by a hyperplane of the form $x_j=\mu_j$. Let $I'=\{i_j\in[3d]\;|\;j\in[d]\setminus[r]\}$. without loss of generality, we assume that $3d\not\in I'$. Then for each $j\in \{r+1,\dots, d\}$ and for any $B,\hat{B}\in\cF_{3d}$
\begin{equation}
    \pi_j(B)\cap\pi_j(\hat{B})\neq\emptyset.
\label{eq:last_fam_proj_1p}\end{equation}

Now suppose $\cF=\big(\bigcup_{i\in [3d]\setminus[p']}\cF_i\big)\cup\{B_1,\dots,B_{p'}\}$ and for any $B\in\cF$, suppose $B'=\prod_{j\in [r]}\pi_j(B)$ and for any $i\in [3d]\setminus[p']$, suppose $\cF'_i=\{B'\;|\;B\in\cF_i\}$ and $\cF'=\{B'\;|\;B\in\cF\}$. Also suppose that $D'=\prod_{j\in [r]}[\beta_{j_1},\alpha_{j_2}]$. Then using the similar arguments as in Case 1  on $\cF'$ and $D'$, we can show that there is a diagonally opposite pair $(\lambda,\lambda')$ of $D'$ and for each $i\in [3d-1]\setminus[p']$ there exist $\tilde{B}'_i\in\cF'_i$ such that
\begin{equation}
    (\lambda,\lambda')\text{ pierce }\cB',
\label{eq:proj_2pierceable}\end{equation} where $\cB'=\{B'_1,\dots,B'_{p'}, \tilde{B}'_{p'+1},\dots,\tilde{B}'_{3d-1}\}\cup\cF'_{3d}$.

Suppose $\cB=\{B\in\cF\;|\;B'\in\cB'\}$ and take any $B_a,B_b\in\cB$ and $j\in [d]\setminus[r]$. If $B_a,B_b\in\cF_{3d}$, then by Equation~\eqref{eq:last_fam_proj_1p}, $\pi_j(B_a)\cap\pi_j(B_b)\neq\emptyset.$ Otherwise, $B_a, B_b$ come from two different families and then by Condition~\tc{red}{(iii)}\remove{\ref{Condition (3)}(\Soumi{reference is not correct})}, $\pi_j(B_a)\cap\pi_j(B_b)\neq\emptyset.$ So $\cB_{j} = \left\{ \pi_j(B)\in\RR\;|\;B\in\cB \right\}$ is a family of pairwise intersecting closed intervals and so by Helly's theorem there exists $\nu_j\in\RR$ that pierces $\cB_j$. Hence the hyperplane $x_j=\nu_j$ pierces the family $\cB$, for each $j\in [d]\setminus[r]$. So by Equation~\eqref{eq:proj_2pierceable}, $\cB$ is pierceable by $(L,M)$, where $L=(\lambda_1,\dots,\lambda_r,\nu_{r+1},\dots,\nu_d)$ and $M=(\lambda'_1,\dots,\lambda'_r,\nu_{r+1},\dots,\nu_d)$.

\color{black}

\section{Extremal examples: Proof of Theorem~\ref{Extremal constructions}}\label{sec: extremal constructions}


In this section, we will prove Theorem~\ref{Extremal constructions}, that is, for all $d \in \mathbb{N}$ we have 
$H_c(d,2)>3d-1$.

    




\begin{theorem}[Restatement of Theorem~\ref{Extremal constructions}]
\label{th:restatement-lowerbound}
For every $d\in\mathbb{N}$, there exist non-empty families $\mathcal{F}_{1}, \, \dots, \, \mathcal{F}_{3d-1}$ 
of \remove{axis-parallel }boxes in $\R^d$ such that 
\begin{description}
    \item[(i)] 
        every colorful $(3d-1)$-tuple is $2$-pierceable, and 
    
    \item[(ii)]
        for each $i \in [3d-1]$, $\mathcal{F}_{i}$ is not $2$-pierceable.
\end{description}
The above result together with Theorem~\ref{thm:restatement_strong-h(d,2)}, implies that for all $d \in \mathbb{N}$, we have $H_{c}(d,2) = 3d$.
\end{theorem}

\begin{figure}[ht]
    \centering
    \includegraphics[height= 12.0cm, width= 12.0cm]{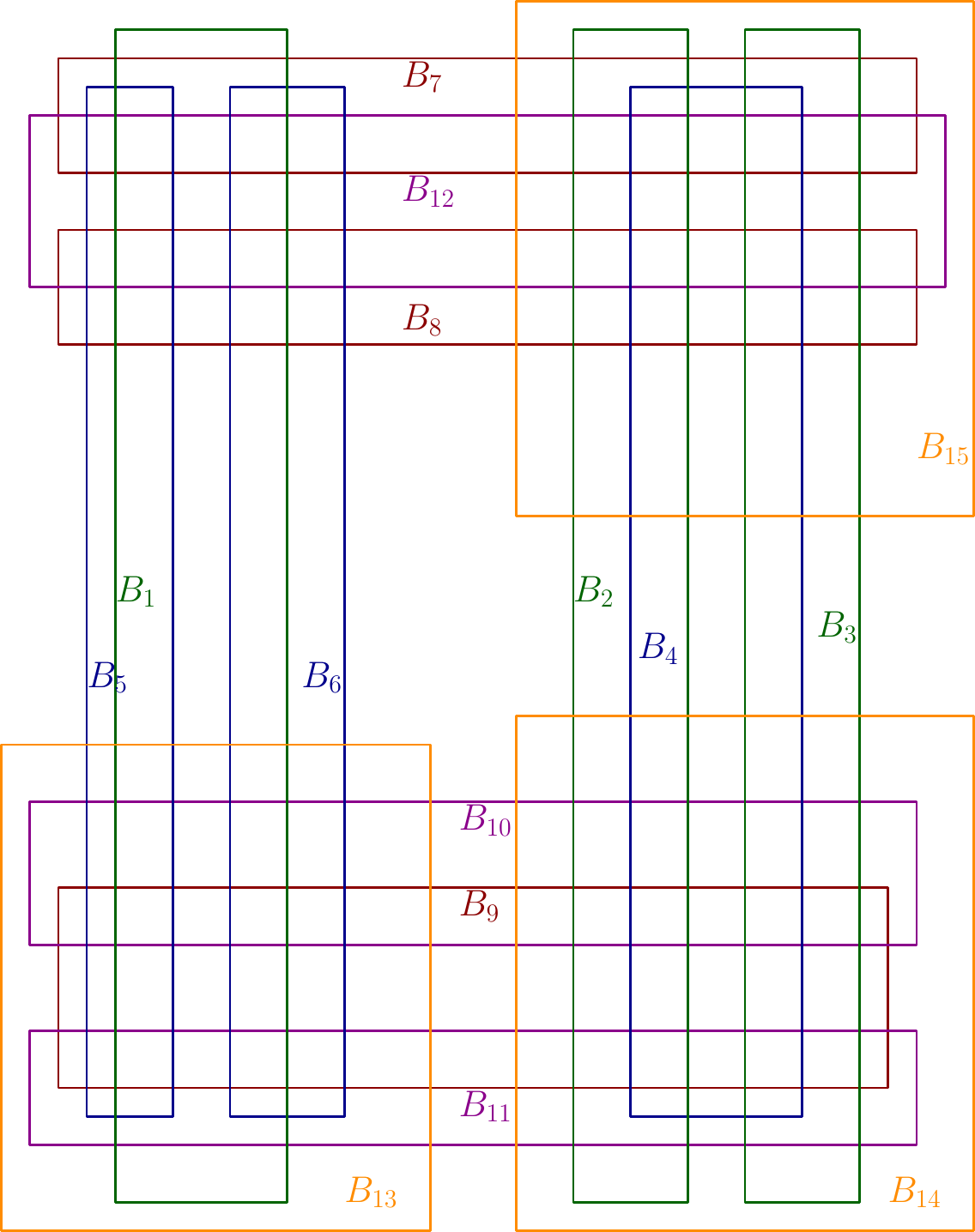}
    \caption{In the figure, for each $i \in \{1, \, 2, \, 3, \, 4, \, 5\}$, $\mathcal{F}_i=\{B_{3i},B_{3i-1},B_{3i-2}\}$.}
    \label{fig:lower_bound}
\end{figure}

Before going into the proof of the above theorem, let us first consider the families $\mathcal{F}_{1}, \, \dots, \, \mathcal{F}_{5}$ of rectangles in $\R^{2}$ given in Figure~\ref{fig:lower_bound}. Observe that any colorful $5$-tuple is $2$-pierceable but none of the families are $2$-pierceable. The main difficulty in the proof of Theorem~\ref{th:restatement-lowerbound} is to construct familes $\mathcal{F}_{1}, \dots, \mathcal{F}_{3d-1}$, for $d \geq 4$, satisfying conditions (i) and (ii).   
With this in mind, we shall now give the details of the construction.

\begin{table}[h]
\begin{center}
    \begin{tabular}{|c|c||c|c|}
    \hline
    \quad \quad $B_{2j-1}$ \quad \quad & \quad \quad $B_{2j}$ \quad \quad & \quad \quad ~~$\alpha_j$ \quad \quad & \quad \quad $\beta_j$ \quad \quad \\
    \hline\hline
         \quad \quad $B_{2j-1,1}$ \quad \quad & \quad \quad $B_{2j,1}$ \quad \quad & \quad \quad $-2$ \quad \quad & \quad \quad $2$ \quad \quad \\
    \hline
        \quad \quad $B_{2j-1,1}$ \quad \quad & \quad \quad $B_{2j,2}$ \quad \quad & \quad \quad $-2$ \quad \quad & \quad \quad $1$ \quad \quad \\ 
    \hline
        \quad \quad $B_{2j-1,1}$ \quad \quad & \quad \quad $B_{2j,3}$ \quad \quad & \quad \quad $-2$ \quad \quad & \quad \quad $2$ \quad \quad \\
    \hline
        \quad \quad $B_{2j-1,2}$ \quad \quad & \quad \quad $B_{2j,1}$ \quad \quad & \quad \quad ~~$0$ \quad \quad & \quad \quad $2$ \quad \quad \\ 
    \hline
        \quad \quad $B_{2j-1,2}$ \quad \quad & \quad \quad $B_{2j,2}$ \quad \quad & \quad \quad ~~$0$ \quad \quad & \quad \quad $1$ \quad \quad \\ 
    \hline
        \quad \quad $B_{2j-1,2}$ \quad \quad & \quad \quad $B_{2j,3}$ \quad \quad & \quad \quad $-1$ \quad \quad & \quad \quad $1$ \quad \quad \\
    \hline
        \quad \quad $B_{2j-1,3}$ \quad \quad & \quad \quad $B_{2j,1}$ \quad \quad & \quad \quad ~~$0$ \quad \quad & \quad \quad $2$ \quad \quad \\ 
    \hline
        \quad \quad $B_{2j-1,3}$ \quad \quad & \quad \quad $B_{2j,2}$ \quad \quad & \quad \quad ~~$0$ \quad \quad & \quad \quad $1.5$ \quad \quad \\ 
    \hline
        \quad \quad $B_{2j-1,3}$ \quad \quad & \quad \quad $B_{2j,3}$ \quad \quad & \quad \quad $-2$ \quad \quad & \quad \quad $2$ \quad \quad \\
    \hline
   \end{tabular}

\caption{Table for $\alpha_j$ and $\beta_j$.}
\label{fig:box} 
\end{center}

\end{table}

\begin{proof}[Proof of Theorem~\ref{th:restatement-lowerbound}]
The main idea of the construction of families $\mathcal{F}_{1}, \dots, \mathcal{F}_{3d-1}$ satisfying conditions (i)
and (ii) of Theorem~\ref{th:restatement-lowerbound} is the following: using colorful $2d$-tuples from the first $2d$-families we will identify a box $D$ such that every colorful $(3d-1)$-tuple will be hit by at least one pair of diagonally opposite vertices of $D$. The construction becomes interesting for $d\geq 4$, as in these cases after identifying $D$, we have to construct the remaining more than two families keeping in mind that every colorful $(d-1)$-tuple from these families are hit by at least one pair of diagonally opposite vertices of $D$.

For each $i\in [d]$, let
$\mathcal{F}_{2i-1}= \left\{B_{2i-1,1}, B_{2i-1,2}, B_{2i-1,3}\right\},\, \mbox{and}\, \mathcal{F}_{2i}=\{B_{2i,1}, B_{2i,2}, B_{2i,3}\},$ 
where
\begin{itemize}
    \item $B_{2i-1,1}=\Big\{(x_{1},\dots, x_d)\in \mathbb{R}^{d} \; \big| \; -4\leq x_i\leq -2,\,\mbox{and}\,\forall j \neq i,\, -4\leq x_j\leq 4 \Big\}$
    
    \item $B_{2i-1,2} = \Big\{(x_{1},\dots, x_d)\in\mathbb{R}^{d} \; \big| \; -1\leq x_i\leq 0,\, \mbox{and}\, \forall j\neq i,
    -4 \leq x_{j} \leq 4 \Big\}$
    
    \item $B_{2i-1,3} = \Big\{(x_{1},\dots, x_d)\in\mathbb{R}^{d} \; \big| \; 1.25\leq x_i\leq 3,\, \mbox{and}\, \forall j\neq i, -4\leq x_j\leq 4 \Big\}$

    \item $B_{2i,1} = \Big\{(x_{1},\dots, x_d)\in\mathbb{R}^{d} \; \big| \; 2\leq x_i\leq 4,\,\mbox{and}\, \forall j\neq i, -4\leq x_j\leq 4 \Big\}$
    
    \item $B_{2i,2} = \Big\{(x_{1},\dots, x_d)\in\mathbb{R}^{d}\; \big| \; 1\leq x_i\leq 1.5,\,\mbox{and}\, \forall j\neq i, -4\leq x_j\leq 4 \Big\}$
    
    \item $B_{2i,3} = \Big\{(x_{1},\dots, x_d)\in\mathbb{R}^{d} \; \big| \; -2.5\leq x_i\leq -1.5,\,\mbox{and}\, \forall j\neq i, -4\leq x_j\leq 4\Big\}$
\end{itemize}
\noindent
For each $i\in [d-1]$, let $\mathcal{F}_{2d+i} = \left\{B_{2d+i,1}, B_{2d+i,2}, B_{2d+i,3}\right\}$ where

\begin{itemize}
    \item $B_{2d+i,1} = \Big\{(x_{1},\dots, x_d)\in\mathbb{R}^d\; \big| \; -5\leq x_1\leq 0.25, -5\leq x_{d-(i-1)}\leq 0.25,\,\mbox{and}\, \forall j\not\in \{1, d-(i-1)\},\, -5\leq x_j\leq 5 \Big\}$
    
    \item $B_{2d+i,2} = \Big\{(x_{1},\dots, x_d)\in\mathbb{R}^d \; \big| \; -5\leq x_1\leq 0.25, 0.75\leq x_{d-(i-1)}\leq 5,\,\mbox{and}\,\forall j\not\in \{1, d-(i-1)\}, -5\leq x_j\leq 5 \Big\}$
    
    \item $B_{2d+i,3} = \Big\{(x_{1},\dots, x_d)\in\mathbb{R}^d \; \big| \; 0.75\leq x_1\leq 5, 0.75\leq x_{d-(i-1)}\leq 5,\,\mbox{and}\,\forall j\not\in \{1, d-(i-1)\}, -5\leq x_j\leq 5 \Big\}$
    
\end{itemize}

\noindent Observe that for all $k\in [3d-1]$ and $B,\, B'\in \mathcal{F}_k$, we have $B\cap B'=\emptyset$. Hence, $\mathcal{F}_k$ is not $2$-pierceable.

Now let $(B_1, B_2, B_3,\dots, B_{3d-1})$ be a colorful $(3d-1)$-tuple, where $B_{k}\in \mathcal{F}_k$ for all $k \in [3d-1]$.
For each $j\in[d]$, depending upon $B_{2j-1}, B_{2j}$, we shall give two real numbers $\alpha_j, \beta_j$ in Table~\ref{fig:box} such that for each $j\in [d]$ and $k\in [3d-1]$,  $\pi_j(B_{k})\cap\{\alpha_j,\beta_j\}\neq\emptyset$ and every diagonally opposite pair of vertices of the box $D=\{(x_1, x_2,\dots, x_d)\in\mathbb{R}^d \; | \; \alpha_j\leq x_j\leq \beta_j, \forall j\in[d]\}$ hit $\{B_1, B_2,\dots, B_{2d}\}$.

\begin{table}[h]
    \begin{center}
   
    \begin{tabular}{|c|c|c|}\hline
        \quad $B_{2d+k}$ \quad & \quad $x_{d-(k-1)}$ \quad & \quad $y_{d-(k-1)}$ \quad \\ \hline\hline
        \quad $B_{2d+k,1}$ \quad & \quad $\alpha_k$ \quad & \quad $\beta_k$ \quad \\ \hline
        \quad $B_{2d+k,2}$ \quad & \quad $\beta_k$ \quad & \quad $\alpha_k$ \quad \\ \hline
        \quad $B_{2d+k,3}$ \quad & \quad $\alpha_k$ \quad & \quad $\beta_k$ \quad \\ \hline
   \end{tabular}

   \caption{Table gives the co-ordinates of $X$ and $Y$ except the first co-ordinate}
\label{fig:diag_pts} 
    \end{center}
    
\end{table}

Now let $X,Y$ be two diagonally opposite vertices of the box $D$, where $X=(x_1,x_2,\dots,x_d)$ and $Y=(y_1, y_2,\dots,y_d)$ such that $x_1=\alpha_1$ and $y_1=\beta_1$. In Table~\ref{fig:diag_pts} we define the other co-ordinates $x_{d-(k-1)}$ and $y_{d-(k-1)}$ of $X$ and $Y$ respectively depending upon  $B_{2d+k}$ for all $k\in[d-1]$.
Observe that $\{X,Y\}$ hits the family of boxes $\{B_1, B_2,\dots, B_{3d-1}\}$. This completes the proof of the theorem.
\end{proof}

\section{Conclusion}
\label{sec:conclusion}

Given a convex body $\mathcal{K}$ in $\mathbb{R}^{d}$, let $h_{\mathcal{K}}(n)$ denote the smallest integer such that any family of translates of $\mathcal{K}$ in $\mathbb{R}^{d}$ is $n$-pierceable if and only if every subset of at most $h_{\mathcal{K}}(n)$ translates is $n$-pierceable. Danzer and Gr\"unbaum~\cite{DBLP:journals/combinatorica/DanzerG82} showed the existence of a centrally symmetric convex body $\mathcal{K}$ in $\mathbb{R}^{2}$ for which $h_{\mathcal{K}}(2) = \aleph_0$, and when $\mathcal{K}$ is a convex polytope they conjectured the following result. 

\begin{conjecture}[Danzer and Gr\"unbaum~\cite{DBLP:journals/combinatorica/DanzerG82}]
    \label{Danzer-Grunbaum-82}
    For all convex polytope $\mathcal{K}$ in $\mathbb{R}^{d}$, $h_{\mathcal{K}}(2) < \aleph_0$.
\end{conjecture}

Gr\"unbaum~\cite{Grunbaum59} conjectured that if $\mathcal{F}$ is a family of translates of a compact, centrally symmetric convex set in $\mathbb{R}^{2}$ such that every pair of translates in the family intersects, then $\mathcal{F}$ is $3$-pierceable. Nearly forty years later, Karasev~\cite{Karasev00} resolved this conjecture. 
Dol\'nikov\footnote{At an Oberwolfach workshop in Discrete Geometry in 2011.} conjectured the following colorful version of Karasev's result.

\begin{conjecture}[Dol\'nikov 2011]
\label{conj: Dolnikov}
Let $K$ be a convex set in the plane, and let $F_1, F_2, F_3$ be families of translates of $K$.
If any two translates of \( K \) from different families intersect, then there exists an index \( i \) such that \( F_i \) is $3$-pierceable.
\end{conjecture}

\noindent
So far, the conjecture has been verified for triangles and for {\em centrally symmetric} convex bodies; see~\cite{DBLP:journals/dm/Jeronimo-Castro15}. Further investigation of Conjecture~\ref{Danzer-Grunbaum-82} and Conjecture~\ref{conj: Dolnikov} is an interesting research direction.

\section*{Acknowledgements}

Sourav Chakraborty acknowledges partial support from the Department of Science and Technology (DST), Government of India, through grant TPN-104427. Arijit Ghosh acknowledges partial support from the Science and Engineering Research Board (SERB), Government of India, through the MATRICS grant MTR/2023/001527, and from the Department of Science and Technology (DST), Government of India, through grant TPN-104427.

\bibliographystyle{alpha}
\bibliography{ref}

\end{document}